# Profile Popularity in a Business-oriented Online Social Network

Thorsten Strufe
P2P Networking Group, TU Darmstadt
strufe[at]cs.tu-darmstadt.de

*Abstract*—Analysing Online Social Networks (OSN), voluntarily maintained and automatically exploitable databases of electronic personal information, promises a wealth of insight into their users' behavior, interest, and utilization of these currently predominant services on the Internet. To understand popularity in OSN, we monitored a large sample of profiles from a highly popular network for three months, and analysed the relation between profile properties and their impression frequency. Evaluating the data indicates a strong relation between both the number of accepted contacts and the diligence of updating contacts versus the frequency of requests for a profile. Counter intuitively, the overall activity, gender, as well as participation span of users have no remarkable impact on their profile's popularity.

## I. Introduction

Social networking services (SNS) contain a wealth of information. Users voluntarily feed self-descriptory details into the Online Social Network (OSN) and their utilization behavior is completely observable by the OSN provider. Analysing this complex data facilitates understanding of the psychological and sociological properties of online social networks and their users. Understanding how users navigate online social networks, e.g., offers insight into how people browse the presented profiles. It consequently allows for the identification of relations between selected profile characteristics and their request frequency. We will call this frequency *popularity* ($\pi$) for the reminder of this paper, which we will measure as the number of impressions of a profile $P$ (the number of times it has been viewed by another user) per time $t$ (or: $\pi_t(P)$).

Identifying key properties of users and their profiles, which allow for the prediction of their popularity, however, is not only interesting for social scientists. Especially system designers and developers of social networking services may capitalize on the extracted knowledge. In case of centralized, server based systems, like facebook[1], LinkedIn[1], or twitter[1], the user experience may be enhanced. Being able to predict the popularity helps suggesting timing constraints for different profiles. Designers hence are able to decide, which profiles need to be presented with very low delay under normal circumstances. The possibility to identify properties of profiles that may tolerate higher response times is an additional, beneficial effect. It additionally yields information, which profiles may more frequently tolerate temporary inavailabilities without causing a significant deterioration of the service experience.

Knowledge on the popularity of profiles may prove being even more valuable for the development of decentralized architectures [1], [2], for the parametrization of this entirely deviating approach to providing social networking services.

Rough, intuitive beliefs about the profile popularity are abundant. Unverified rumours frequently have it that profiles of women are more often visited than profiles of men, that providing a profile picture will drastically increase the number of impressions of a profile, or that the unfortunate possession of a last name late in the alphabet, will inevitably condemn a user's profile to eternal lack of popularity.

This work aims at exploring these conceptions. It analyzes data from a large, central European OSN with a main focus on professional- and business centered utilization. The main contributions are

- to corroborate the fact that correlations between properties of profiles and their popularity do exist, as well as
- to expose, which properties of profiles have a considerable effect on their popularity, and which properties have less, or no effect at all.

A large set of profiles from the selected OSN are monitored for this purpose, and traces that were gathered over a period of three months in this course have been analysed. Mann Whitney U tests have been used to identify significant differences in the popularity of profiles of different groups, and correlations between parametric variables and the profiles' popularity have been estimated, using Pearson's rank-sum test.

The results indicate that the gender of users actually does not have an impact on profile popularity, while providing a profile picture does. The alphabetic order of users' last names, or the past membership duration in general do not significantly correlate with profile popularity, whereas the membership in interest groups, the frequency of profile updates, and the number of contacts in the OSN actually do.

The remaining paper is divided into seven sections. Following an overview of the state of the art (section II), the monitored OSN and its properties are described (III), and a description, of how the analysed data has been gathered, is given (IV). Section V describes the processing of the collected data, before section VI describes how it has been refined for analysis and checked for validity. Subsequently, section VII presents the statistical analysis and results, followed by section VIII, which concludes this work by summarizing the results and denoting our current, and future work.

[1]http://www.[facebook|linkedin|twitter].com

## II. PREVIOUS INTERPRETATIONS AND ANALYSES OF OSN

Examining Online Social Networks with lack of access to the server logs and databases can generally be done using one of three different approaches: using *surveys* and *interviews*, by *traffic logging*, and by conducting *crawls* and monitoring the data through the web interface of the OSN directly (Schneider et al. [3] give an insightful and very comprehensive overview of the related work).

Surveys and interviews [4], [5], while allowing for the very detailed and fine grained examination of opinions, properties, and behavior, are naturally bound to very small sample sizes for reasons of resource constraints. Aiming at more general insights, a more scalable approach is needed for this work.

Logging and analysing Internet traffic is another approach to understanding utilization of, and user behavior in OSN. These studies [3], [6] offer first statistics on the different topics and services used inside the analysed OSN, as well as their users' bandwidth consumption, and session characteristics. This approach, while potentially covering a large subset of profiles of an OSN, can only achieve local subsampling of requests that traverse the tapped ISP. It thus can only stochastically gather information, but performed on a short term is unable to give complete insight into the popularity of requested profiles.

Crawling accesses a subset of the registered profiles and hence represents a subsampling of the examined OSN. While unable to cover the whole social network (at least for large OSN), it gathers the complete public data and properties of the visited profiles. Previous studies that harnessed crawlers [7], [8] have generally aimed at understanding structural properties of OSN graphs. They mainly conclude that the network graph exhibits power-law and small world properties. Viswanath et al. [9], similarly to our approach, have conducted multiple crawls of selected profiles in recent work. They have focused on monitoring the interaction between users via leaving wall posts, though. This allows for inferences with respect to direct interaction between selected users. It may give slight hints on the popularity of users and profiles, as well, but draws an incomplete picture to this end. Lampe et al. [10] is probably most related to our work. In their study, the authors crawled over 30.000 profiles in facebook and discovered how providing certain fields in profiles influenced the expected size of their friend list. In this paper we go a step further. On top of analyzing static properties of profiles in an OSN and correlating them to the number of accepted contacts, we monitor the dynamic changes to profiles over a large period of time. Using this data, we derive, which utilization behavior of users and which properties of their profiles have an influence on their own profile's popularity.

## III. EXAMINED OSN AND ITS PROPERTIES

A major, central European OSN has been selected for the purpose of this study. This OSN contains over 8 million registered users and is focused on professional and business-oriented networking, much alike the internationally well accepted LinkedIn. The contained profiles consequently are almost entirely completed and maintained very well.

Commonly registered information in the OSN can be grouped into five categories: *identifying information, curriculum vitae, interests, personal contacts,* and *messaging* (cmp. Table I). Additional details, like the registration date, a user activity estimation, and total number of profile impressions are automatically integrated into the profiles by the OSN provider, without any participation of the user. The activity metric is not very fine grained, nor does it support any conclusion on session length or exact times of logins. It still does give indication of the broad frequency of a user's OSN utilization. The gender of the users is not explicitly stated in this OSN, but the first names are freely accessible on the profile.

| Available Profile Information in the analysed OSN | |
| --- | --- |
| Identifying information | Name, photo, birthday, address information (physical and electronic) |
| Curriculum vitae | Current and previous employers, job titles and descriptions, education, attended universities and schools, known languages |
| Interests | Personal interest, professional interest, wants, haves, subscribed interest groups |
| Personal contacts | Contact list |
| Messaging | Guest book entries, recommendations |

Table I
USER MAINTAINED DETAILS IN PROFILES

A distinctive feature of the monitored OSN is the fact, that all changes and updates to any registered profile detail are indicated to the accepted contacts of the user. To this end, the last five updates to the profiles of a user's contacts are shown on the main page of the OSN, unlike in facebook, where the "news feed" is assembled from arbitrary updates of friends.

## IV. DATA COLLECTION

Access to profiles and the full user behavior for scientific purposes is rather difficult, since, for obvious commercial and privacy reasons, OSN providers are keen on protecting and highly unwilling to disclose them. Some information still can be gathered by accessing the publicly available profiles through the web interface of the SNS directly.

The main purpose of participating in an OSN being to publish information about oneself, the users quite willingly share detailed information about themselves, usually including the list of other users they have contacted. This data is rarely protected by privacy measures and usually shared even with the broad public without any access restrictions.

To retrieve data for analyses in this study, crawlers keep constantly monitoring a selection of profiles in the OSN. This selection, a subset of all profiles, needs to be large enough to achieve reliable results, but limited for reasons of resource restrictions on both the OSN provider's and the researcher's side. To choose the selection, several independent random walks of different length through the social network have been performed. Profiles at presumed edges of the network in presumed large mutual distance have manually been selected as origins of the random walks, to foster a diverse sampling.

Three of the random walks, spanning 25.000, 5.000, and 1.643 steps respectively, have finally been selected as the sample set. In this selection it was assured that neither the traversed profiles of these three random walks, nor their contact lists contained any overlapping users. An interesting, yet, in retrospection rather unsurprising observation has been that all random walks quickly converged to the core of the network (the majority of users is from German speaking countries), which they subsequently never left again. A quite high number of independent random walks consequently had to be conducted, in order to sample independent subsets of profiles for the study, and a total of over 2 million distinct profiles have been visited in this process.

| Collected data of the monitored profiles | |
| --- | --- |
| Identifying information | Name, login, photo provided (boolean), gender (as derived through third party web-page) |
| Curriculum vitae | Current type of employment, universities attended, spoken languages |
| Interests | Interests, subscribed interest groups (including the number of members of the group, number of messages in group, supported languages) |
| Personal contacts | Number of contacts (complete contact list is monitored infrequently) |
| OSN specific data | Registration date, number of profiles impressions (in total since registration), estimated activity |
| Monitoring details | Time stamp of each monitoring visit |

Table II
REGULARLY COLLECTED PROFILE DETAILS

Ever since their selection, the 31.643 profiles are monitored regularly by bots, which were implemented especially for this task. Since only selected data was needed for the analysis, and images are distinguished by name and hence do not need to be retrieved, the bots record only extracts of data during each monitoring run (cmp. Table II). Since an initial hypothesis had been that the gender would have strong influence on the popularity of profiles, this detail was needed. However, the data available from the OSN did not include any explicit information on the gender of the profile owners. To still reliably determine the gender of the users in an automated fashion, an interface to a major website containing international first names[2] has been created. Through this interface the bot requested information about the first name of each profile once. This process was successful with the exception of three first names, which internationally are used for both men and women. The gender of the three respective profiles could easily be derived by checking the profiles manually.

## V. Data refinement

A subsample from the complete data had to be extracted for analysis. It includes the all profile traces over a period of three months (Nov 2009 – Jan 2010). The bots visited the majority of profiles over 400 times, and at least twice per day during this time span (with the exception of a two week period, in which the monitoring was performed slightly less regularly for reasons of unreliable connectivity to the service). The overall size of log files of extracted information for the span of three months is just over 1.3 GB.

[2]http://www.mein-vorname.com

Processing the raw monitored data was needed to allow for meaningful analyses. Considering the different registration dates, and the fact that the profile properties are only available for the time span of monitoring, the number of profile impressions has been converted to the *popularity* of profile impressions per hour: $\pi_h(P)$. Possessing the collection of traces over time, a fine-grained analysis of selected intervals might give interesting insight into detailed behavior of users. Yet, throughout this study the overall average of profile impressions per hour over the complete monitoring span of three months has been used.

| Processed variables for analysis per profile | |
| --- | --- |
| Non-parametric | First name, last name, gender, availabe photo, login handle |
| Parametric | Membership duration, final number of friends, final number of profile impressions, average activity (given by provider), $\pi_h(P)$, duration of monitoring, final no. of subscribed groups, final no. of provided interests, final no. of attended universities, final no. of stated spoken languages, no. changes to employment status, no. changes to interests, no. changes to friendlist, overall no. profile alterations |

Table III
ANALZED DETAILS AFTER PROCESSING

The activity estimation from the OSN turned out to be of limited value, since it exhibits some sublinear decrease on inactivity but some seemingly linear increase on activity of the users. With traces of all changes to the essential details of each profile, a second measure of activity fortunately could be established. Tracking each change to a profile, the *alteration frequency*, representing the number of detail updates per day, has been calculated for all profiles as an alternative activity measure. This metric was determined for changes of the employment status, changes to the list of interests, and changes in the number of accepted contacts, as well as the sum of all three. In contrast to the activity as served by the provider, this metric of course does not reflect passive or pure consumption behavior of users at all. It can only display active profile maintenance, or alterations to the contact list. Then again, all influences on it are known and it can explicitly be analysed, unlike the rather imprecise average activity from the profiles.

In consequence, a set of 19 variables, five of which were non-parametric, and 14 parametric (cmp. Table III), were available for each profile.

## VI. Sanitizing the Monitored Data

The 31.643 profiles consisted of 21.436 male ($m$), and 10.207 female ($f$) users. Before examining it, some profiles from this large sample were removed for different reasons: Three of the randomly selected profiles were the official profiles of celebrities and represented extreme outliers (each with more than three times the profile impressions per hour than any of the remaining profiles). 29 profiles have been removed, since they had been deregistered during the monitoring span. Further profiles were removed due to the fact, that they seemed to be abandoned (the popularity was too low and the activity of the related user could not reliably be derived).

The remaining sample (cmp. Table IV) consisted of 17.450 profiles of male and 7.824 profiles of female users, adding up to 25.274 profiles in total. The provider states a ratio of 34% out of the overall user base being female, which is quite close to the 31% measured in the sample.

In order to better understand the characteristics of the sample, it has been examined closer with respect to the different groups contained. 23.766 profiles included a picture ($m : 16.387 / f : 7.379$), whereas in 1508 profiles the picture has been left blank ($m : 1063 / f : 445$). With respect to the number of accepted contacts, the minimum in the sample was 5 and the maximum 12.332, the degree distribution follows a power law (cmp. Fig. 1, inset). The highest number of subscribed interest groups measured was 511, with a median of 3. A large set of profiles (5387) are not subscribed to any groups at all.

Normalizing the profile popularity to average profile impressions per hour $\pi_h$, the profiles in the sample ranged from 0.002 $\pi_h$ and 6.56 $\pi_h$. Examining the popularity distribution, binned to profile impressions per week, we expected to observe a Zipf-distribution, considering that it reflects user preferences, which commonly are expected to be Zipf-distributed. The fitting yielded an $s \approx 1.4, D = -771$, an insufficient result (cmp. Fig. 2). Fitting to a log-Normal distribution led to much more satisfying results with $\mu \approx 1.96$ and $\sigma \approx 0.865$.

Testing the gender against the average activity and our alteration frequency, it turned out that male users are generally more active than female users. The results match, when comparing the distributions of the average activity ($Mdn = 87$ ($m$) $> Mdn = 85$ ($f$), $p < 0.001$) as well as the alteration frequency ($Mdn = 0.04$ ($m$) $> Mdn = 0.0384$ ($f$), $p < 0.001$). Considering our alteration frequency, this difference is significantly supported ($p < 0.001$) for all contained variables, the change in status, interests, and contact lists.

Testing the fact, if a picture is provided, against the average activity and our alteration frequency did not yield any surprising results: the users who upload a picture into their profile are significantly more active than users who do not. The average activity distributions differ by a large distance ($Mdn = 87$ (picture provided, $pic$) $> Mdn = 47$ (no picture provided, $npic$, $p < 0.001$), as well as the distributions in alteration frequency ($Mdn = 0.04$ ($pic$) $> Mdn = 0.02$ ($npic$), $p < 0.001$).

To further validate the sample, the frequency of first names and last names has been fitted to test for power law and Zipf distributions, respectively. Fitting the degrees to a power law distribution yielded in an alpha of $\alpha = 2.91$ and fitting the ranks of name frequencies to Zipf-distributions yielded in an exponent $s$ of 3.14 for last names and an $s$ of 1.67 for first names (fitting to exponential functions led to $k = -3$ for last

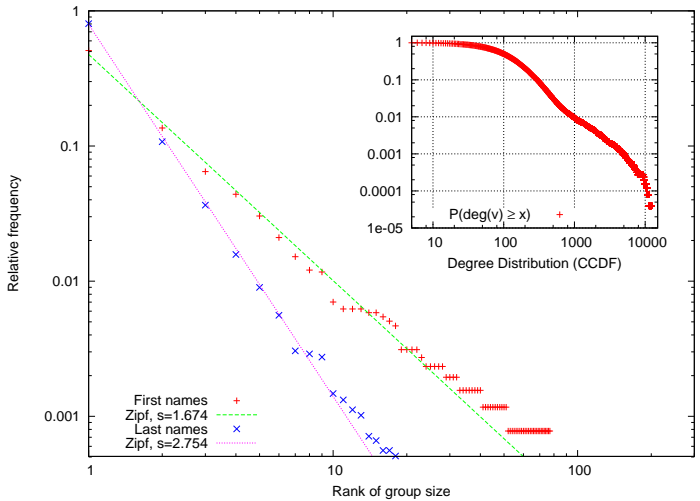

Figure 1. Distributions (relative frequency over ranks of groupsizes) of first- and last names in the sample, with fitted Zipf distributions (log-log), and CCDF of the degree distribution over contacts $deg(P)$ (inset)

names and $k = -1.85$ for first names, cmp. Fig. 1). These results are in-line with previously measured values, and the sample in consequence has been accepted as the basis for further analyses [8], [11].

## VII. POPULARITY OF PROFILES

In order to determine the relations between the available profile properties and the popularity $\pi_h(P)$ of profiles, we conducted a test for differences in distributions of the non-parametric measures, first, and a correlational study using the parametric data, after.

### A. Non-parametric Tests

Analysing non-parametric relationships between classes of users and their corresponding profile popularity was performed using Mann-Whitney U (Wilcoxon rank-sum) tests. The main tested hypotheses were the identity of the profile popularity of male vs. female users and of profiles including a picture vs. profiles without.

The first hypothesis was that the user's gender would have an impact on the popularity of their profiles. This hypothesis was not supported by the data, and profiles of male ($Mdn = 0.039$) and female ($Mdn = 0.041$) users did not display significantly different popularity ($ns$, all results cmp. Table V).

Testing the impact of providing a profile photo yielded the expected results, which supported the hypothesis that profiles including a photo ($Mdn = 0.043$) were significantly more often requested than profiles without a photo ($Mdn = 0.016$, $p < 0.001$). The distance was estimated to be $d = 0.024$, or roughly one profile impression every two days. Inspecting the influence of the gender on the subsamples yielded no surprises. Significance and estimated distances where almost identical when testing the subsamples of both male and female users'

| Sample description of monitored profiles for popularity analyses | | | |
|---|---|---|---|
| Total profiles | 25.274 | 17.450 $male$, 7.824 $female$ | |
| Profile photo | 23.766 ($m$: 16.387, $f$: 7.379) | none: 1508 ($m$: 1063, $f$: 445) | |
| *Variable* | *Min* | *Median* | *Max* |
| Activity | 0.01 ($m,f$: 0.01) | 86.2 ($m$: 86.9 ,$f$: 84.6) | 100 ($m,f$: 100) |
| Profile alterations (per day) | 0 ($m,f$: 0) | 0.04 ($m,f$: 0.04) | 1.20 ($m$: 1.18) |
| Degree (contacts) | 5 ($m,f$: 5) | 101 ($m$: 104, $f$:93) | 12.332 ($m$: 11.225) |
| Groups subscribed | 0 ($m,f$: 0) | 3 ($m$: 3,$f$: 2) | 511 ($f$: 107) |
| Popularity (PI/h) | 0.002 ($m,f$: 0.002) | 0.04 ($m,f$: 0.04) | 6.56 ($m$: 5.48) |

Table IV
SAMPLE DESCRIPTION OF THE POPULARITY EXPERIMENT

profiles, when comparing the respective profiles including pictures against those without.

*B. Correlational Tests*

A correlational study has been performed following the non-parametric tests. Tests for a normal distribution failed for all monitored variables. Considering the large sample sizes, Pearson's $r$ nonetheless was used to determine potential correlations between parametric properties of the profiles and their corresponding popularity.

Before testing six main hypotheses on the profile popularity, including relations between a user's activity, group membership, profile degree, membership duration, and alphabetic order of last names, it was examined if the membership duration since profile registration had any impact on the degree.

Preferential attachment on mind, the correlation between membership duration and the number of contacts was estimated. Testing $H_0$ that the true correlation between membership duration and the number of contacts was 0 failed ($p < 0.001$), indicating that a correlation indeed exists. Pearson's $r$ was estimated to $r \approx 0.19$, a slight correlation, with $0.18 < p < 0.20$ to a confidence of 95%. Subsequently, the samples were tested for correlations of monitored parameters and $\pi_h(P)$.

Considering the fact that highly active users are expected to interact with other users more and hence are expected to be more visible than others, the average activity has first been studied for a correlation with popularity. Pearson's test was significant, and the hypothesis $H_0$ that there is no correlation between *activity and popularity* was broken ($p < 0.001$), indicating that indeed the activity and popularity are correlated. Pearson's product moment correlation was estimated to be 0.17, representing a slight correlation between activity of the users and the popularity of their profiles. This correlation is slightly, but not significantly higher for male users ($0.16 < r < 0.19$ with $p < 0.025$) than for female users ($0.14 < r < 0.18$ with $p < 0.025$) in the sample.

Comparing this result to the correlation between the alteration frequency and profile popularity yielded surprising results. Again, Pearson's test was significant ($p < 0.001$), and the product moment correlation between the alteration frequency of a profile and its popularity has been estimated to a high 0.62 ($0.61 < r < 0.63$ with $p < 0.025$). This correlation may be caused by the fact that the last five profile changes of contacts are presented to the users after login, and a constant profile alteration consequently increases the chances of a user to be visible with his contacts.

To test, if participation in interest groups had an impact, the correlation between the number of subscribed groups and the profile popularity was tested. The results again refuse a correlation of 0 between subscribed groups and $\pi_h(P)$ and suggest a correlation of $r \approx 0.37$. In this case, the correlation is significantly higher for male users ($0.37 < r < 0.4, p < 0.025$) compared to female users ($0.33 < r < 0.37, p < 0.025$).

It further was studied, if a correlation between the degree of a profile (the number of accepted contacts) and its popularity could be found. The hypothesis that the correlation between *degree and popularity* is 0 fails, again, and the Pearson test indicates a quite high correlation between the two variables of $r \approx 0.75$ ($p < 0.001$). This correlation is significantly higher for profiles of female users ($0.81 < r < 0.83, p < 0.025$) compared to male users ($0.74 < r < 0.75, p < 0.025$). This result is not too surprising, since a higher degree obviously leads to a higher chance of randomly being "browsed upon", and changes to the profile additionally are presented at more login pages of other users, thus leading to a higher visibility.

Knowing about the correlation of membership duration and the number of accepted contacts, the data was analysed for a correlation between the membership duration and the profile popularity, as well. The test indeed indicated a slight correlation between membership duration and the popularity. Yet, at $r \approx 0.11$ ($p < 0.001$) it was quite low.

A final hypothesis reflects the intuitive question, if the user's name has an impact on the popularity of the profile. All profiles being presented in increasing alphabetic order of the last name is the simple rationale behind this conjecture: Users with last names starting with a character early in the alphabet might plainly enjoy a higher visibility due to the fact that other users start browsing contact list pages, which usually contain only ten contacts, without pursuing to the later pages and hence not reaching "late" contacts, at all. The tested correlation is very slightly, but not significantly determined for the whole group of users ($r = -0.01, ns$). Analysing the *rich club* of profiles with the highest popularity, in contrast, leads to an impressive change of results. Considering the 5% profiles with the highest popularity already yields a correlation of $r \approx -0.09$ ($ns, -0.26 < r < 0.08$ with $p < 0.025$). This correlation gets more significant with increasing "exclusivity" of the rich club: Analysing the top 2‰ of profiles the correlation increases to $r \approx -0.22$ ($ns, -0.47 < r < 0.06$ with $p < 0.025$), for the top 1‰ of profiles to $r \approx -0.29$ ($ns$, $-0.62 < r < 0.11$ with $p < 0.025$), and for the top 10 users, it finally increases to $r \approx -0.9, p < 0.001$ ($-0.98 < r < -0.61$ with $p < 0.025$). Considering the very small samples size, this result of course has to be taken with a grain of salt.

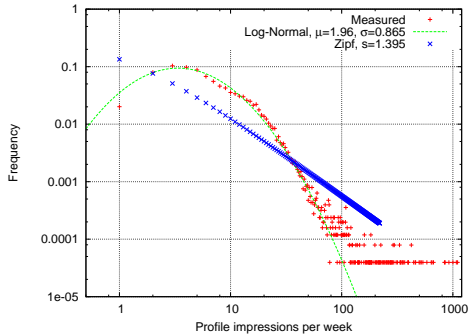

| Similarity and difference of means of *Popularity* $\pi_{v_i}$ (rank sum) | | | |
|---|---|---|---|
| Gender | Male $Mdn = 0.039$ | Female $Mdn = 0.041$ | no significant difference |
| Profile Picture | Picture ($Mdn = 0.043$) > No picture ($Mdn = 0.016$) ($p < 0.001$) | | |
| Correlational observations (Pearson's $r$) | | | |
| Accepted contacts | correlation: $r \approx 0.75$ ($p < 0.001$) | | |
| | Male | $0.74 < r < 0.75$ | ($p < 0.05$) |
| | Female | $0.81 < r < 0.83$ | ($p < 0.05$) |
| Profile alterations | correlation: $r \approx 0.62$ ($p < 0.001$) | | |
| Subscribed groups | correlation: $r \approx 0.37$ ($p < 0.001$) | | |
| | Male | $0.37 < r < 0.4$ | ($p < 0.05$) |
| | Female | $0.33 < r < 0.37$ | ($p < 0.05$) |
| Average activity | correlation: $r \approx 0.17$ ($p < 0.001$) | | |
| | Male (0.18) | Female (0.16) | no significant difference |

Table V
SUMMARY OF THE POPULARITY EXPERIMENT

Figure 2. Popularity distribution with log-Normal fit ($\mu = 1.96, \sigma = 0.865$) and insufficient Zipf-fit (log-log)

Summarizing the results, it becomes clear that gender and last names have no serious impact on the profile popularity, while providing a profile picture, being active in interest groups and the number of accepted contacts clearly influence the popularity in a positive way.

## VIII. SUMMARY AND OUTLOOK

This paper deals with the popularity of profiles in Online Social Networks (OSN). Popularity has been defined as the average number of profile impressions per hour, and it has been analysed on data from a major, central-European OSN with a focus on professional use. 31.643 profiles have been selected by random walks and subsequently monitored for this purpose. A section covering 3 months of monitoring data from the overall set has been extracted, outliers have been removed, and the remaining sample of profiles has been validated.

Non-parametric tests were performed to estimate the differences of popularity between groups of users, especially considering their gender and if a profile picture had been provided. A correlational study was performed on parametric data subsequently, to estimate the impact of profile characteristics.

The results underline that certain properties actually are good indicators for the expected popularity of profiles. They significantly suggest strong correlations between the subscription of groups, the activity of users, especially with profile alterations, and most importantly the number of accepted contacts of a profile with its popularity. Additionally, they suggest that profiles with pictures will be more frequently viewed than profiles without, while no significant difference between the profiles of male vs. female users could be determined.

Following these promising first results, we are aiming at extending our studies. The study analysed traces that were gathered over a period of three months, while the monitoring has been kept active in the meantime. Testing exemplary results of the presented study against the larger data set, which since has expanded to span a good four months of monitoring, does not yield any significant changes. It hence does not seem to be beneficial to extend the monitoring period, since the results are bound to differ only marginally.

So far, the data has solely been tested as a whole, though. The activity and profile popularity have only been calculated in average over the whole period, without taking into account temporary popularity changes, e.g., after extensive profile alterations. We consequently have started to analyse the data over the course of the monitoring time, to identify possible timed dependencies.

Additionally, we have gained access to the complete webserver log files and databases of a regional, facebook-like university OSN, and started to process this wealth of data. It of course promises much more detailed insight into the user behavior and popularity, which we are about to start to investigate.